\documentclass{cpbtex}

\usepackage{bm}
\usepackage{amsmath}
\usepackage{amssymb}
\usepackage{graphicx}
\usepackage{subfigure}
\usepackage{hyperref}
\usepackage{color}
\usepackage{orcidlink}
\usepackage{array}
\usepackage{indentfirst}
\usepackage{orcidlink}
\hypersetup{
	colorlinks=true,
	linkcolor=red,
	citecolor=blue,
}
\newcommand{\hb}[1]{\bar{h}^{(#1)}}
\newcommand{\f}{\frac}
\newcommand{\shift}{\beta}

\newcommand{\ljump}{\big[\!\big[}
\newcommand{\rjump}{\big]\!\big]}
\newcommand{\bracketl}{\big\{\!\big\{}
\newcommand{\bracketr}{\big\}\!\big\}}
\newcommand{\CZ}[1]{{\color{black}{#1}}}

\begin{document}
\title{The point-particle-limit effective-source approach for computing gravitational self-force in the Lorenz gauge}

\author{Chao Zhang(张超)$^{1}$, \ Yungui Gong(龚云贵)$^{1}$\thanks{Corresponding author. E-mail:~gongyungui@nbu.edu.cn}, \ Xuchen Lu(路旭晨)$^{1}$, \\ Wenting Zhou(周文婷)$^{1}$\\
$^{1}${Institute of Fundamental Physics and Quantum Technology,}\\
{School of Physical Science and Technology,}\\
{Ningbo University, Ningbo, Zhejiang 315211, China}
}

\date{\today}
\maketitle

\begin{abstract}
The traditional effective-source method is hampered by complex analytical expressions and the inherent smoothness limit, which incur high computational costs and complicate implementation.
To overcome these limitations, we introduce the point-particle-limit effective source method, which analytically takes the size of the effective source to zero, thereby transforming the problem into a well-defined jump condition of retarded metric field at the particle position governed by the local singular field.
This formulation naturally pairs with a discontinuous Galerkin scheme, whose inherent capacity for accommodating solution discontinuities enables highly accurate enforcement of the jump conditions.
We apply both the traditional and point-particle-limit effective source method to calculate the time-domain gravitational metric perturbation and gravitational self-force in the Lorenz gauge on a point particle in a circular orbit around a Schwarzschild black hole.
\CZ{The method of point-particle-limit effective source method achieves gravitational self-force relative errors below $10^{-4}$ (temporal) and $10^{-3}$ (radial) at $\ell_{\rm max}=15$ for circular Schwarzschild orbits, reduces computational cost by an order of magnitude compared to the traditional approach, and yields energy fluxes that agree with frequency-domain benchmarks.}
The comparison of numerical results shows the excellent advantage of the point-particle-limit effective source method, which validates the \CZ{effectiveness} and efficiency of the point-particle-limit effective source method and thereby establishes a numerical foundation for computing generic geodesic orbits or long-time self-consistent orbital evolution.
\end{abstract}



\section{Introduction}
The gravitational self-force (GSF) problem \cite{Barack:2009ux,Barack:2018yvs, Poisson:2011nh,Mino:1996nk} constitutes a central challenge in the theoretical modeling of extreme-mass-ratio inspirals (EMRIs) \cite{Amaro-Seoane:2007osp, Babak:2017tow}, which are systems comprising a stellar-mass compact object spiraling into a supermassive black hole, with the mass ratio in the range of $10^{-7}-10^{-4}$.
EMRIs represent primary targets for future space-based gravitational wave observatories, including the Laser Interferometer Space Antenna (LISA) \cite{Danzmann:1997hm,Audley:2017drz},
TianQin \cite{Luo:2015ght} and Taiji \cite{Hu:2017mde}.
These observations will provide unprecedented opportunities to map the spacetime geometry of supermassive black holes with high precision and to test general relativity in the strong-field regime \cite{Amaro-Seoane:2007osp,Babak:2017tow,Berry:2019wgg,Fan:2020zhy,Zi:2021pdp,Destounis:2020kss,Destounis:2021mqv,Destounis:2021rko,Cardoso:2021wlq,Hou:2024rgo,Zhang:2024ogc}.
To fully realize the scientific potential, theoretical waveform templates must achieve a phase accuracy of order $0.1$ radian over the course of the multi-year inspiral, which requires a precise description of the small body's motion beyond the geodesic approximation \cite{TianQin:2020hid,Ruan:2018tsw,LISA:2022yao,LISA:2022kgy}.

In EMRIs, perturbation theory in the small mass-ratio allows the small object to move along a geodesic of the larger body's background spacetime at zeroth order, with finite-mass corrections such as radiation reaction incorporated order by order via the GSF $-$ the backreaction of the particle's own perturbative field.
The GSF includes both dissipative effects that drive orbital decay and conservative effects that modify orbital frequencies like pericenter precession.
However, a fundamental difficulty arises from the point-particle idealization that the gravitational perturbation diverges on the worldline, making the bare GSF formally infinite.
Extracting a finite, physically meaningful GSF therefore requires a systematic regularization procedure to subtract the singular part of the field, which is associated with the particle's instantaneous presence and does not affect its motion.

Self-force phenomena first emerged in electrodynamics through the work of DeWitt and Brehme \cite{DeWitt:1960fc}, with foundational studies also carried out for scalar fields as a simpler analog model that captures many of the key features of the gravitational problem \cite{Quinn:2000wa,Barack:2000eh,Nakano:2001kw,Barack:2002mha,Diaz-Rivera:2004nim}.
The gravitational case was later addressed by Mino, Sasaki $\&$ Tanaka \cite{Mino:1996nk} and independently by Quinn $\&$ Wald \cite{Quinn:1996am}, leading to the MiSaTaQuWa formulation, in which the small body departs from geodesic motion due to its own backreaction.
An equivalent perspective was provided by Detweiler and Whiting \cite{Detweiler:2002mi}, who described the dynamics as geodesic motion in a suitably perturbed spacetime.
Practical computations became feasible with the development of the mode-sum regularization scheme, largely advanced by Barack and Ori \cite{Barack:1999wf,Barack:2001bw,Barack:2001gx,Barack:2002bt,Barack:2002mh}, first applied to scalar and electromagnetic fields, then extended to the gravitational field \cite{Hughes:2005qb,Barack:2005nr,Barack:2007tm,Barack:2007we,Akcay:2010dx,Akcay:2013wfa}.
In the mode-sum approach, the retarded field perturbation is decomposed into spherical harmonic modes, each mode is regularized by subtracting analytic terms derived from the local singular field, and the summed contributions converge to a finite self-force.
This framework has been successfully applied to scalar, electromagnetic, and gravitational self-forces across a range of configurations, including circular and eccentric orbits in Schwarzschild spacetime, and more recently to orbits in Kerr spacetime \cite{vandeMeent:2017bcc,Torres:2020fye,Heffernan:2021olv}. 
Unfortunately, the mode-sum regularization scheme, successful at first-order GSF calculation, becomes markedly less effective when extended to the second-order GSF problem.
At the second order, the retarded field contains divergent terms that scale as even powers of the distance from the particle.
This stems from the fact that the second-order field involves quadratic combinations of first-order quantities.
Unlike the odd-power case, the angular decomposition of such even-power singularities yields individual modes that diverge logarithmically as the worldline is approached \cite{Wardell:2015kea}.

To calculate the second-order GSF, the effective source approach was proposed independently by Barack and Golbourn \cite{Barack:2007jh}, and by Vega and Detweiler \cite{Vega:2007mc}.
The central idea of the effective source method is to construct a regularized effective source such that the resulting residual field is smooth on the worldline and directly yields the GSF without the need for mode-by-mode subtraction.
This approach offers improved convergence properties and avoids the explicit handling of divergent quantities. 
The effective source approach was first developed and applied in the scalar-field case, serving as a crucial testing ground before its extension to tensor field.
Early work demonstrated its feasibility in the time domain for scalar self-forces on circular orbits \cite{Vega:2011wf}.
Building on this foundation, the framework was later extended to handle eccentric trajectories, with new techniques such as the method of extended effective sources introduced to overcome the slow convergence of Fourier sums for non-circular orbits \cite{Leather:2023dzj}.
A significant milestone followed with the first fully analytic construction of effective sources for arbitrary geodesic motion around Schwarzschild black hole \cite{Wardell:2011gb}, providing a critical foundation for self-consistent orbital evolution in the scalar-field setting \cite{Heffernan:2017cad,Diener:2011cc}.
With the scalar case well established, the effective source method was subsequently adapted to GSF calculations, beginning with frequency-domain implementations for circular orbits in Schwarzschild spacetime \cite{Wardell:2015ada}.
More recently, the effective source framework has become the workhorse of second-order GSF computations, which underlies recent post-adiabatic waveform models \cite{Wardell:2021fyy,Spiers:2023mor,Albertini:2024rrs,Wei:2025lva,Upton:2025bja,Warburton:2021kwk,Miller:2023ers}.

The traditional effective source (TES) approach is limited by complex analytical expressions and the inherent smoothness constraint, leading to high computational costs and implementation complexity. 
In this work, we introduce the point-particle-limit effective source (PPLES) method, which analytically takes the source size to zero, reducing the problem to a well-defined jump condition on the retarded metric field at the particle location, governed by the local singular field. 
This formulation naturally pairs with a discontinuous Galerkin (DG) scheme, whose inherent capacity for accommodating solution discontinuities enables highly accurate enforcement of the jump conditions.
Unlike continuous finite element methods requiring smooth sources, the DG scheme permits discontinuities across element boundaries that the particle position is treated as an interface with jump conditions enforced weakly through numerical fluxes-achieving high-order accuracy essential for GSF precision in gravitational-wave applications.
Specifically, we consider a compact object on a circular orbit in Schwarzschild spacetime, which serves as the simplest yet physically relevant testbed for validating new computational methods.
This work lays the groundwork for extensions to eccentric orbits, Kerr backgrounds, and second-order GSF within a unified time-domain approach.

This paper is organized as follows. 
Section 2 reviews Lorenz-gauge gravitational perturbation theory and the effective source method.
Section 3 presents the method of the TES method with the worldtube scheme and the PPLES method with the DG scheme.
Section 4 describes the numerical implementation details, including coordinate transformation, discretization formulation, and numerical fluxes for the DG scheme.
Section 5 presents numerical validation results for circular Schwarzschild orbits, verifying the accuracy and efficiency of the PPLES method.
Sec. 6 is devoted to conclusions and discussions.

\section{The perturbation in Lorenz gauge}
\label{sec2}
In the perturbation framework, the Einstein field equations are
\begin{equation}
  	G_{\mu\nu}[g_{\mu\nu}+h_{\mu\nu}] = 8\pi T_{\mu\nu},
\end{equation}
where $g_{\mu\nu}$ and $h_{\mu\nu}$ are the background spacetime and the metric perturbation, $G_{\mu\nu}$ and $T_{\mu\nu}$ represent the Einstein tensor and the matter stress-energy tensor, respectively.
Imposing the Lorenz gauge condition
\begin{equation}
\nabla_\mu \bar{h}^{\mu\nu}=0,
\end{equation}
the Einstein equations in the first-order metric perturbation can simplify into
\begin{equation}\label{lEQ}
\square\bar{h}_{\mu\nu} + 2 R^{\rho\;\;\sigma}_{\;\;\mu\;\;\nu}\bar{h}_{\rho\sigma} = -16\pi T_{\mu\nu}\, ,
\end{equation}
where $\nabla_\mu$ is the covariant derivative with respect to the background metric, $\bar{h}_{\mu\nu}$ is the trace-reversed perturbation,
\begin{equation}
	\bar{h}_{\mu\nu} \equiv h_{\mu\nu} - \frac{1}{2}g_{\mu\nu}h,
\end{equation}
the trace $h=g_{\mu\nu}h^{\mu\nu}$,
$\square\equiv\nabla_\lambda\nabla^\lambda$ and $R^{\rho\;\;\sigma}_{\;\;\mu\;\;\nu}$ is the Riemann tensor of the background metric.
In the modeling of EMRI systems, the compact object is described by a point particle, resulting in the stress-energy tensor
\begin{equation}\label{eq:energytensor}
	T^{\mu\nu} = m_p\int^\infty_{-\infty} \frac{u^\mu\, u^\nu}{\sqrt{-\det(g)}} \delta^4(x^\mu - x_p^\mu)\, d\tau ,
\end{equation}
where $m_p$ and $u^\mu$ are the mass and four-velocity of the point-particle, $\det(g)$ and $\tau$ are the background metric determinant and the proper time along the particle's worldline, respectively.
The numerical calculation of gravitational self-force in EMRI systems is fundamentally challenging due to the singular nature of point-particle sources at the particle position.
To overcome these difficulties, the effective source method offers an innovative alternative by replacing traditional regularization techniques with a reformulation of the problem.
Following the effective-source approach \cite{Wardell:2015ada}, we can decompose the retarded metric perturbation into singular and regular components,
\begin{equation}\label{eq:hsr}
\bar{h}^{\rm ret}_{\mu\nu}=\bar{h}^{R}_{\mu\nu}+\bar{h}^{S}_{\mu\nu},
\end{equation}
where the regular part $\bar{h}^{R}_{\mu\nu}$
is finite and physically meaningful for numerical computations of gravitational self-force, while the singular part $\bar{h}^{S}_{\mu\nu}$
captures the divergent structure analytically.
Inserting Eq. \eqref{eq:hsr} into Eq. \eqref{lEQ}, the linearized Einstein equations become 
\begin{equation}\label{eff1}
\square\bar{h}^R_{\mu\nu} + 2 R^{\rho\;\;\sigma}_{\;\;\mu\;\;\nu}\bar{h}^R_{\rho\sigma}=S_{\mu\nu}^{\rm eff},
\end{equation}
where the effective source
\begin{equation}
S_{\mu\nu}^{\rm eff} = -16\pi T_{\mu\nu} -\square\bar{h}^S_{\mu\nu} - 2 R^{\rho\;\;\sigma}_{\;\;\mu\;\;\nu}\bar{h}^S_{\rho\sigma}.
\end{equation}

The line element in the background geometry is given by
\begin{equation}
ds^2=-f dt^2+f^{-1}dr^2+r^2d\theta^2+r^2\sin^2\theta d\varphi^2,
\end{equation}
where $f\equiv 1-2/r$ and $(t, r, \theta, \varphi)$ are the standard Schwarzschild coordinates.
Owing to the spherical symmetry of the Schwarzschild background geometry, the metric perturbation can be separable into tensorial spherical harmonics (Barack-Lousto-Sago modes) \cite{Barack:2005nr,Barack:2007tm}
\begin{equation}
\bar{h}_{\mu\nu}=\frac{m_p}{r}\sum_{\ell=0}^{\infty}\sum_{m=-\ell}^{\ell}\sum_{i=1}^{10}a_{\ell}^{(i)}\bar{h}^{(i)}_{\ell m}(r,t)Y_{\mu\nu}^{(i)\ell m}(\theta,\varphi;r),
\end{equation}
where the tensorial-harmonic basis $Y_{\mu\nu}^{(i)\ell m}$ and the normalization factors $a_{\ell}^{(i)}$ are given in  Appendix A.
After extremely tedious computation, the field equations reduce to the coupled set of two-dimensional hyperbolic equations (omitting the indices $\ell m$ unless needed for clarity)
\begin{equation}\label{eq2d}
\begin{split}
\square_{2d}\overline{h}^{(i)}+4\mathcal{M}^{(i)}_{\;(j)}\overline{h}^{(j)}=S_{\rm eff}^{(i)}\quad(i=1,\ldots,10),   \\
\square_{2d}\equiv\frac{\partial}{\partial t^2}-\frac{\partial}{\partial x^2}+V(r),\qquad
V(r)\equiv\frac{f}{r^2}\left[\ell(\ell+1)+\frac{2}{r}\right],
\end{split}
\end{equation}
where the tortoise radial coordinate is defined as
\begin{equation}
x=r+2\ln\left[\frac{r}{2}-1\right],    
\end{equation}
the terms ${\cal M}^{(i)}_{\;(j)}\bar h^{(j)}$ are given in Appendix B and the effective source $S_{\rm eff}^{(i)}$ has be derived in Ref. \cite{Zhang:2025eqz}.

\section{Method}\label{sec3}
The timelike geodesics of the Schwarzschild spacetime on the equatorial plane $\theta_p=\pi/2$ are described by the equations
\begin{equation}
\label{orbite}
\begin{split}
   dt/d\tau &= \frac{E}{f(r_p)},\\
(dr_p/d\tau)^2 &= E^2-f(r_p)\left(1+\frac{L^2}{r_p^2}\right),\\
d\phi_p/d\tau &=\frac{L}{r_p^2},
\end{split}
\end{equation}
where the constants $E$ and $L$ are the energy and angular momentum per unit mass, respectively.
To describe the geodesic orbit, it is useful to make the substitution
\begin{equation}
    r_{p}(\chi)=\frac{p}{1+e\cos\chi}.
\end{equation}
The energy $E$ and angular momentum $L$ are
\begin{equation}
\label{EL}
    \begin{split}
E^2&=\frac{(p-2-2e)(p-2+2e)}{p(p-3-e^2)},\\
L^2&=\frac{p^2}{p-3-e^2}.
    \end{split}
\end{equation}
Integrating Eq.~\eqref{orbite}, we obtain \cite{Cutler:1994pb}
\begin{equation}
  \begin{split}
t(\chi)&=p^2(p-2-2e)^{1/2}(p-2+2e)^{1/2}\\
&\times \int_0^\chi d\chi'(p-2-2e\cos\chi')^{-1}(1+e\cos\chi')^{-2}\\
&\times (p-6-2e\cos\chi')^{-1/2},
  \end{split}
\end{equation}
and
\begin{equation}
\phi_{p}(\chi)=p^{1/2}\int_0^\chi\frac{d\chi'}{(p-6-2e\cos\chi')^{1/2}}.
\end{equation}
In particular, we focus on circular orbits in the equatorial plane of a Schwarzschild black hole to validate the \CZ{effectiveness} and efficiency of our method.

\subsection{Traditional Effective Source Method}
The puncture function $\bar{h}^{(i)}_P$ is constructed by truncating the series expansion of the Detweiler-Whiting singular field $\bar{h}^{(i)}_S$ at second order distance, and the effective source is obtained by applying the relevant wave operators to the puncture field \cite{Vega:2007mc,Barack:2007jh,Vega:2011wf,Zhang:2025eqz};
however, the global behavior of the effective source makes it unsuitable for numerical implementation, as it generally diverges in the large-distance limit.
To address this issue, Vega and collaborators introduce a windowing function that multiplies the puncture, smoothly suppressing the effective source far from the worldline \cite{Vega:2009qb,Vega:2011wf,Wardell:2011gb,Diener:2011cc,Vega:2007mc,Vega:2009qb}.
In contrast, Barack and collaborators adopt a worldtube approach, in which interior and exterior perturbative solutions are matched within a worldtube, seamlessly connecting the solution obtained from an effective source inside the tube to the homogeneous vacuum solution outside \cite{Barack:2007jh,Barack:2007we,Dolan:2010mt,Dolan:2011dx,Dolan:2012jg}.
Both the worldtube and window-function schemes have been successfully demonstrated to be effective regularization techniques, confirming that either can be reliably employed to compute the gravitational metric perturbation, and the worldtube scheme is adopted in this work to solve the field equations numerically.

In the $1+1$D domain, the worldtube $\mathcal{T}$ is defined to be a region of finite extent in radius surrounding the worldline.
Outside the worldtube, we evolve the vacuum field equations for the retarded metric perturbation $\bar{h}^{(i)}_{\rm ret}$.
Inside the worldtube, we evolve the sourced field equations for the regular metric perturbation $\bar{h}^{(i)}_{R}$.
Across the boundary of the worldtube, the retarded and regular metric perturbations are matched via the puncture field.
Our evolution field equations consist of 
\begin{align}
\square_{2d}\overline{h}^{(i)}_{\rm ret}+4\mathcal{M}^{(i)}_{\;(j)}\overline{h}^{(j)}_{\rm ret}=0, \qquad\qquad& \rm{outside}\quad \mathcal{T},\\
\square_{2d}\overline{h}^{(i)}_{R}+4\mathcal{M}^{(i)}_{\;(j)}\overline{h}^{(j)}_{R}=S_{\rm eff}^{(i)},  \qquad\qquad& \rm{inside}\quad \mathcal{T},\\
\overline{h}^{(i)}_{R}=\overline{h}^{(i)}_{\rm ret}-\overline{h}^{(i)}_{P},  \qquad\qquad& \rm{across}\quad \mathcal{T}.
\end{align}

\subsection{Point-Particle-Limit Effective Source Method}
In practice, the expression for the effective source in the worldtube method is computationally expensive and introduces additional numerical errors when evaluating the metric at the boundary of the effective source region.
Motivated by these limitations, we seek to relax the constraints on the size of the effective source.
Using the effective source approach, the regular metric perturbation $\bar{h}^{(i)}_R$ is demonstrated to be continuous across the particle location, as are its time and radial derivatives
\begin{equation}
\begin{split}
\lim_{r \to r_p^+} \bar{h}^{(i)}_R &= \lim_{r \to r_p^-} \bar{h}^{(i)}_R, \\
\lim_{r \to r_p^+} \partial_r \bar{h}^{(i)}_R &= \lim_{r \to r_p^-} \partial_r \bar{h}^{(i)}_R, \\
\lim_{r \to r_p^+} \partial_t \bar{h}^{(i)}_R &= \lim_{r \to r_p^-} \partial_t \bar{h}^{(i)}_R.
\end{split}
\end{equation}
By the matching conditions stated above, we obtain the following relations
\begin{equation}
\begin{split}
\lim_{r \to r_p^\pm} \bar{h}^{(i)}_R &= \lim_{r \to r_p^\pm} \bar{h}_{\rm ret}^{(i)} - \lim_{r \to r_p^\pm} \bar{h}_{P}^{(i)}, \\
\lim_{r \to r_p^\pm} \partial_r \bar{h}^{(i)}_R &= \lim_{r \to r_p^\pm} \partial_r \bar{h}_{\rm ret}^{(i)} - \lim_{r \to r_p^\pm} \partial_r \bar{h}_{P}^{(i)}, \\
\lim_{r \to r_p^\pm} \partial_t \bar{h}^{(i)}_R &= \lim_{r \to r_p^\pm} \partial_t \bar{h}_{\rm ret}^{(i)} - \lim_{r \to r_p^\pm} \partial_t \bar{h}_{P}^{(i)}.
\end{split}
\end{equation}
The jump of retarded field $\bar{h}_{\rm ret}^{(i)}$ between the right-hand and left-hand limits can be derived analytically.
In the limit $r \to r_p^\pm$, we obtain a much simpler expression for this jump compared to the full effective source.
As a concrete example, consider the component $h^{(1)\ell m}_{\rm ret}$ for the specific case of $\ell=2$ and $m=2$.
Its analytical expression takes the form
\begin{equation}
\left[\left[\bar{h}^{(1)}_{\rm ret}\right]\right]=\lim_{r \to r_p^+}  \bar{h}^{(1)}_{\rm ret}-\lim_{r \to r_p^-}  \bar{h}^{(1)}_{\rm ret} =0,   
\end{equation}
\begin{equation}\label{deltadh1dr}
\left[\left[\partial_r\bar{h}^{(1)}_{\rm ret}\right]\right]=\lim_{r \to r_p^+} \partial_r \bar{h}^{(1)}_{\rm ret}-\lim_{r \to r_p^-} \partial_r \bar{h}^{(1)}_{\rm ret} =-\frac{2 \sqrt{30 \pi }  r_p^2E  \left(E^2+(u^r)^2\right)e^{-2 i \phi_p}}{(r_p-2) \left(L^2+r_p^2\right)},   
\end{equation}
\begin{equation}\label{deltadh1dt}
\left[\left[\partial_t\bar{h}^{(1)}_{\rm ret}\right]\right]=\lim_{r \to r_p^+} \partial_t \bar{h}^{(1)}_{\rm ret}-\lim_{r \to r_p^-} \partial_t \bar{h}^{(1)}_{\rm ret} =\frac{u^r(r_p-2)}{r_p E}\frac{2 \sqrt{30 \pi }  r_p^2E  \left(E^2+(u^r)^2\right)e^{-2 i \phi_p}}{(r_p-2) \left(L^2+r_p^2\right)},
\end{equation}
where the expressions for other components and spherical indices $(\ell, m)$ can be found in Ref.~\cite{Zhang:2025eqz}.
Moreover, the jumps in the radial and time derivatives satisfy the identity
\begin{equation}
\left[\left[\partial_t \bar{h}^{(i)}_{\rm ret}\right]\right] = \partial_t \left[\left[\bar{h}^{(i)}_{\rm ret}\right]\right] -\frac{d\,r_p}{dt} \left[\left[\partial_r \bar{h}^{(i)}_{\rm ret}\right]\right],
\end{equation}
which has also been noted in earlier works \cite{Field:2009kk, DaSilva:2024yea}. 
This indicates that the jumps in the field at the particle position are not merely mathematical artifacts but are inherently tied to the structure of the singular field and effective source.
Once the jump of the retarded field $\bar{h}^{(i)}_{\rm ret}$ at the particle is known, it can be incorporated into a DG scheme to accurately resolve the field equation
\begin{equation}
\square_{2d}\overline{h}^{(i)}_{\rm ret}+4\mathcal{M}^{(i)}_{\;(j)}\overline{h}^{(j)}_{\rm ret}=0 \,,
\end{equation}
together with the jump conditions $\left[\left[\bar{h}^{(i)}_{\rm ret}\right]\right]$, $\left[\left[\partial_t\bar{h}^{(i)}_{\rm ret}\right]\right]$ and $\left[\left[\partial_r\bar{h}^{(i)}_{\rm ret}\right]\right]$.

\section{Numerical implementation details}\label{sec4}
\subsection{Coordinate Transformation}
To effectively resolve moving discontinuities within the DG framework, particularly those arising from complex particle motion like geodesics, the optimal strategy is to adopt a coordinate system where the discontinuity is stationary.
The fundamental challenge is that a moving discontinuity, such as one tied to a particle on a geodesic, traverses the computational domain, making it difficult to track accurately with a fixed mesh and complicating flux integration.
This problem can be addressed by a coordinate transformation technique motivated by the problem of a constrained moving particle \cite{Field:2009kk}. 
For such a particle in the domain $a < x_p(t) < b$, the transformation from $(t,x)$ to $(\lambda,\xi)$ is given by
\begin{align}
t & = \lambda, \\
x & = a
+ \frac{x_p(\lambda) - a}{\xi_p -a}(\xi-a)
+ \frac{(b-x_p(\lambda))(\xi_p-a)-(x_p(\lambda)-a)(b-\xi_p)}{(\xi_p-a)(b-\xi_p)(b-a)}
(\xi-a)(\xi-\xi_p),
\label{eq:xofxi}
\end{align}
where $x_p(\lambda)$ is explicitly time-dependent, while $\xi_p$ is time-independent, ensuring that the particle's position is fixed at $\xi_p$.
In this spirit, applying a transformation designed to immobilize the moving discontinuity at $\xi_p$ ensures the governing equations are properly transformed, thereby streamlining the subsequent DG discretization.
Taking the derivatives of Eq. \eqref{eq:xofxi} gives
\begin{equation}
\begin{split}
x_{\lambda}&=\frac{\partial x}{\partial\lambda} =
\frac{(\xi-a)(b-\xi) x_p'(\lambda)}{(\xi_p-a)(b-\xi_p)} ,
\\
x_\xi&=\frac{\partial x}{\partial\xi}  =
\frac{(2\xi-\xi_p - a)(\xi_p-x_p(\lambda))
+(x_p(\lambda)-a)(b-\xi_p)}{(\xi_p-a)(b-\xi_p)},
\\
x_{\xi\xi}&=\frac{\partial^2 x}{\partial\xi^2}  =
\frac{2(\xi_p- x_p(\lambda))}{(\xi_p-a)(b-\xi_p)},
\end{split}
\end{equation}
where the shift vector is 
\begin{equation}\label{eq:explicit_shiftvector}
\beta^\xi
= \frac{\partial x/\partial \lambda}{\partial x/\partial \xi}
= \frac{(\xi-a)(b-\xi) x_p'(\lambda)
}{(2\xi-\xi_p - a)(\xi_p-x_p(\lambda))
+(x_p(\lambda)-a)(b-\xi_p)}.
\end{equation} 
Its differentiation is
\begin{equation}
\frac{\partial \beta^\xi}{\partial\xi} = 
\frac{(A\xi^2 + B \xi + C) x_p'(\lambda)}{
\big[(2\xi-\xi_p - a)(\xi_p-x_p(\lambda))
+(x_p(\lambda)-a)(b-\xi_p)\big]^2},
\end{equation}
where $A = 2(x_p(\lambda)-\xi_p)$, $B = 2 (a b + \xi_p^2 - (a + b) 
x_p(\lambda))$, and $C = (a^2 + b^2) x_p(\lambda)
- a (b - \xi_p)^2 - b (a^2 + \xi_p^2)$. 

We use the letter $\Psi^{(i)}(\lambda,\xi)$ to denote the field $\bar{h}^{(i)}(t,r)$ in the new coordinates.
To convert the original second-order wave equation into a first-order system in the $(\lambda,\xi)$ coordinates, we introduce auxiliary variables
\begin{equation}\label{variables}
\begin{split}
    \Phi^{(i)} &= \partial_\xi\Psi^{(i)}=(\partial x/\partial\xi)\partial_x\Psi^{(i)}=(\partial x/\partial\xi)f(r)\partial_r\Psi^{(i)},\\
    \Pi^{(i)}  &=-\partial_t\Psi^{(i)}=-\partial_\lambda\Psi^{(i)}+\beta^\xi \Phi^{(i)}.
\end{split}    
\end{equation}
The resulting first-order system in the $(\lambda,\xi)$ coordinates is given by
\begin{equation}\label{firstEQ}
    \begin{split}
        \partial_\lambda\Psi^{(i)} & = \beta^\xi \Phi^{(i)} - \Pi^{(i)}, \\
\partial_\lambda\Pi^{(i)} & = \beta^\xi \partial_\xi \Pi^{(i)}
-(\partial x/\partial\xi)^{-1}\partial_\xi [(\partial x/\partial\xi)^{-1}\Phi^{(i)}]
+ V(r)\Psi^{(i)} +4\mathcal{M}^{(i)}_{\;(j)}\Psi^{(i)},
\\
\partial_\lambda \Phi^{(i)} & = \partial_\xi (\beta^\xi \Phi^{(i)}-\Pi^{(i)}),
    \end{split}
\end{equation}
where the cross term $\mathcal{M}^{(i)}_{\;(j)}\Psi^{(i)}$ make different $i$th $(\Psi^{(i)},\Pi^{(i)},\Phi^{(i)})$ coupled with each other, but with no derivative term.

\subsection{Discontinuous Galerkin method}\label{sec:dgscheme}
Following Refs. \cite{Field:2009kk,10.5555/1557392}, the computational domain $\Omega$ is taken as the closed $\xi$-interval $[a,b]$. We partition $\Omega$ into $K>1$ non-overlapping subintervals $\mathsf{D}^k = [a^k,b^k]$, where $a = a^1$, $b = b^K$, and $b^{k-1} = a^k$ for $k = 2,\cdots,K$.
The particle location is assumed to be at $\xi_p = b^{k_p} = a^{k_p+1}$, i.e., the endpoint shared by $\mathsf{D}^{k_p}$ and $\mathsf{D}^{k_p+1}$, with $1 \leq k_p < K$.
Within each subinterval $\mathsf{D}^k$, each component of the system vector $(\Psi^{(i)},\Pi^{(i)},\Phi^{(i)})$ is approximated by a local interpolating polynomial of degree $N$.
In terms of the column vectors
\begin{equation}
\bm{\Psi}^{(i)k}(\lambda) 
= \big[\Psi^{(i)}(\lambda,\xi^k_0),\cdots,
\Psi^{(i)}(\lambda,\xi^k_N)\big]^T,\qquad
\bm{\ell}^k(\xi)
= \left[\ell^k_0(\xi),\cdots,\ell^k_N(\xi)\right]^T,
\end{equation}
the approximation is expressed as
\begin{equation}
\Psi^{(i)k}_h(\lambda,\xi) = 
\Psi^{(i)k}(\lambda)^T 
\ell^k(\xi),
\end{equation}
where $\ell^k(\xi)_j$ is $j$th the Lagrange polynomial for the subinterval $\mathsf{D}^k$.
For each subinterval $\mathsf{D}^k$ and each component of the solution, the local residuals are defined as follows
\begin{subequations}\label{eq:residuals}
\begin{align}
(R_{\Psi^{(i)}})^k_h & = \partial_\lambda \Psi^{(i)k}_h
                 - (\shift^\xi\Phi^{(i)})^k_h + \Pi^{(i)k}_h ,
\\
(R_{\Pi^{(i)}})^k_h & = \partial_\lambda \Pi^{(i)k}_h
- \partial_\xi (\shift^\xi \Pi^{(i)})^k_h
+ (\Pi^{(i)} \partial_\xi \shift^\xi)^k_h
+\partial_\xi (x_\xi^{-2}\Phi^{(i)})^k_h
+\cdots ,
\\
(R_{\Phi^{(i)}})^k_h & = \partial_\lambda \Phi^{(i)k}_h
- \partial_\xi (\shift^\xi \Phi^{(i)})^k_h 
+ \partial_\xi\Pi^{(i)k}_h.
\end{align}
\end{subequations}
The upwind numerical fluxes are obtained
\begin{subequations}\label{eq:NumericalFlux}
\begin{align} 
(f_{\Pi}^k)^* & = 
\bracketl {-\shift^\xi} \Pi_h + x_\xi^{-2}\Phi_h \bracketr 
+ \frac{1}{2}
\ljump x_\xi^{-1}\Pi_h -x_\xi^{-1}\shift^\xi\Phi_h \rjump_\mathrm{num}, 
\\
(f_{\Phi}^k)^* & = 
\bracketl \Pi_h  - \shift^\xi\Phi_h \bracketr 
+ \frac{1}{2}
\ljump x_\xi^{-1}\Phi_h -x_\xi\shift^\xi\Pi_h \rjump_\mathrm{num},
\end{align}
\end{subequations}
where the average of a numerical variable and its numerical jump are
\begin{equation}
\bracketl A \bracketr = \frac{1}{2}(A^+ + A^-),
\qquad
\ljump A \rjump_\mathrm{num} 
= \mathbf{n}^+ A^+ + \mathbf{n}^- A^- ,
\end{equation}
$\mathbf{n}^-$ denotes the outward normal from the local subdomain, and $\mathbf{n}^+=-\mathbf{n}^-$.
For the endpoints $a^{k_p+1}$ and $b^{k_p}$ at the particle position, the jump condition should be applied to the numerical flux.
The average of a numerical variable remains unchanged, and its numerical jump at the particle position should be modified into
\begin{equation}
\ljump \Pi \rjump_\mathrm{num} 
= \mathbf{n}^+ \Pi^+ + \mathbf{n}^- \Pi^-+\mathbf{n}^-\left[\left[\Pi_{\rm S}\right]\right],
\qquad
\ljump \Phi \rjump_\mathrm{num} 
= \mathbf{n}^+ \Phi^+ + \mathbf{n}^- \Phi^- +\mathbf{n}^-\left[\left[\Phi_{\rm S}\right]\right],
\end{equation}
where the jump of $\left[\left[\Pi_{\rm S}\right]\right]$ and $\left[\left[\Phi_{\rm S}\right]\right]$ can be derived by Eqs. \eqref{deltadh1dr} and \eqref{deltadh1dt} and Eq. \eqref{variables}.
At the boundaries of outgoing and ingoing radiation  conditions, the Sommerfeld boundary conditions are
\begin{equation}
(\partial_t\Psi -\partial_x\Psi)(\lambda,a) = 0
\rightarrow \Pi(\lambda ,a)+\Phi(\lambda ,a)/x_\xi(\lambda,a)=0,
\end{equation}
\begin{equation}
(\partial_t\Psi+\partial_x\Psi)(\lambda,b) = 0
\rightarrow \Pi(\lambda ,b)-\Phi(\lambda ,b)/x_\xi(\lambda,b)=0.    
\end{equation}

\section{Numerical Results}\label{sec5}
The metric perturbations are computed using both the TES method and the PPLES method, and a detailed comparison of their numerical behavior is performed. 
While the primary results focus on the quadrupole mode $\ell=2$, the performance of the scheme extends naturally to higher multipoles.
However, within the present formulation, the dipole mode $\ell=1$ and the monopole mode $\ell=0$ suffer from the divergence problem.
Instead, we deal with the two modes using the standard frequency-domain method.
The physical Lorenz-gauge monopole and dipole are constructed from a basis of homogeneous frequency-mode solutions of the underlying ordinary differential equations.
The computational domain is defined with inner and outer boundaries at $a=-100$ and $b=500$, respectively, and the particle is located at $\xi_p = 20$.
A total of $120$ subdomains are employed, each with $N=15$ collocation points, and the system is evolved with a timestep $\Delta\lambda = 0.0005$.
The initial data are set to zero, $\bar{h}^{(i)}=0$.

The metric perturbations are computed using two distinct approaches.
The first is the TES method, implemented via the world-tube construction in the scalar-field evolution framework~\cite{Heffernan:2017cad}, which is available in BHPToolkit~\cite{BHPToolkit}.
The second is the PPLES method, which bypasses the need for an effective source by directly incorporating the analytic jumps of the retarded field.
For a circular orbit with radius $r_p = 10$, the two methods should yield identical regular field results in principle. However, Figure~\ref{Fig1:e0} shows that at the particle location, the regular field obtained from the PPLES method does not match the TES result.
\begin{center}
\refstepcounter{figure}
\includegraphics[width=0.99\textwidth]{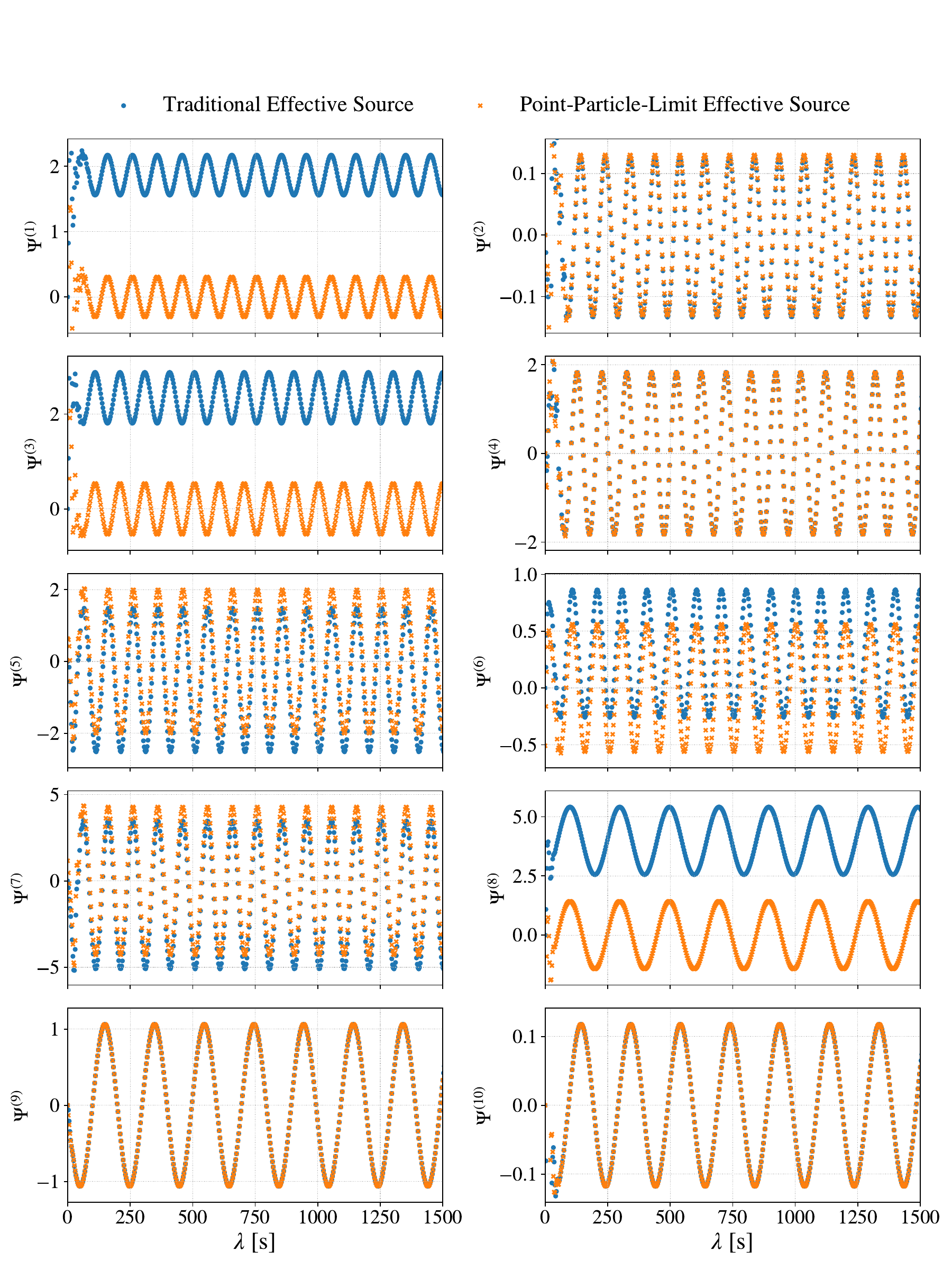}\\[8pt]
\parbox[c]{15.0cm}{\footnotesize{\bf Fig.~\thefigure.} Time evolution of the real components of the regular metric field $\bar{h}^{(i)}_R=\Psi^{(i)}$ at the particle position for a circular orbit around a Schwarzschild black hole. Blue points show results from the TES method, and orange points correspond to the PPLES method. \CZ{Following the left-to-right, top-to-bottom order of the figure, the first seven display the even-parity components $\bar{h}^{(1-7)}_R$ for $(\ell,m)=(2,2)$, and the last three show the odd-parity components $\bar{h}^{(8-10)}_R$ for $(\ell,m)=(2,1)$.}}
\label{Fig1:e0}
\end{center}
The discrepancy is systematic, as for TES, the averaged regularized field $\bar{h}^{(i)}_{R}$ does not remain at zero but instead oscillates about a non-zero offset, which is inconsistent with the expected behavior of the Lorenz-gauge wave equation for circular orbits.
In contrast, the PPLES results oscillate symmetrically around zero, as required.
The difference between the two solutions manifests predominantly as a constant shift.
Taking a time slice at $\lambda = 1500$ s, Figure~\ref{Fig2:e0} shows additional discrepancies outside the world-tube. Specifically, for TES, discrepancies appear between the retarded field outside the effective source region and the regular field inside the source region; for PPLES, discrepancies are observed in the retarded field outside the particle location.
\begin{center}
\refstepcounter{figure}
\includegraphics[width=1.0\textwidth]{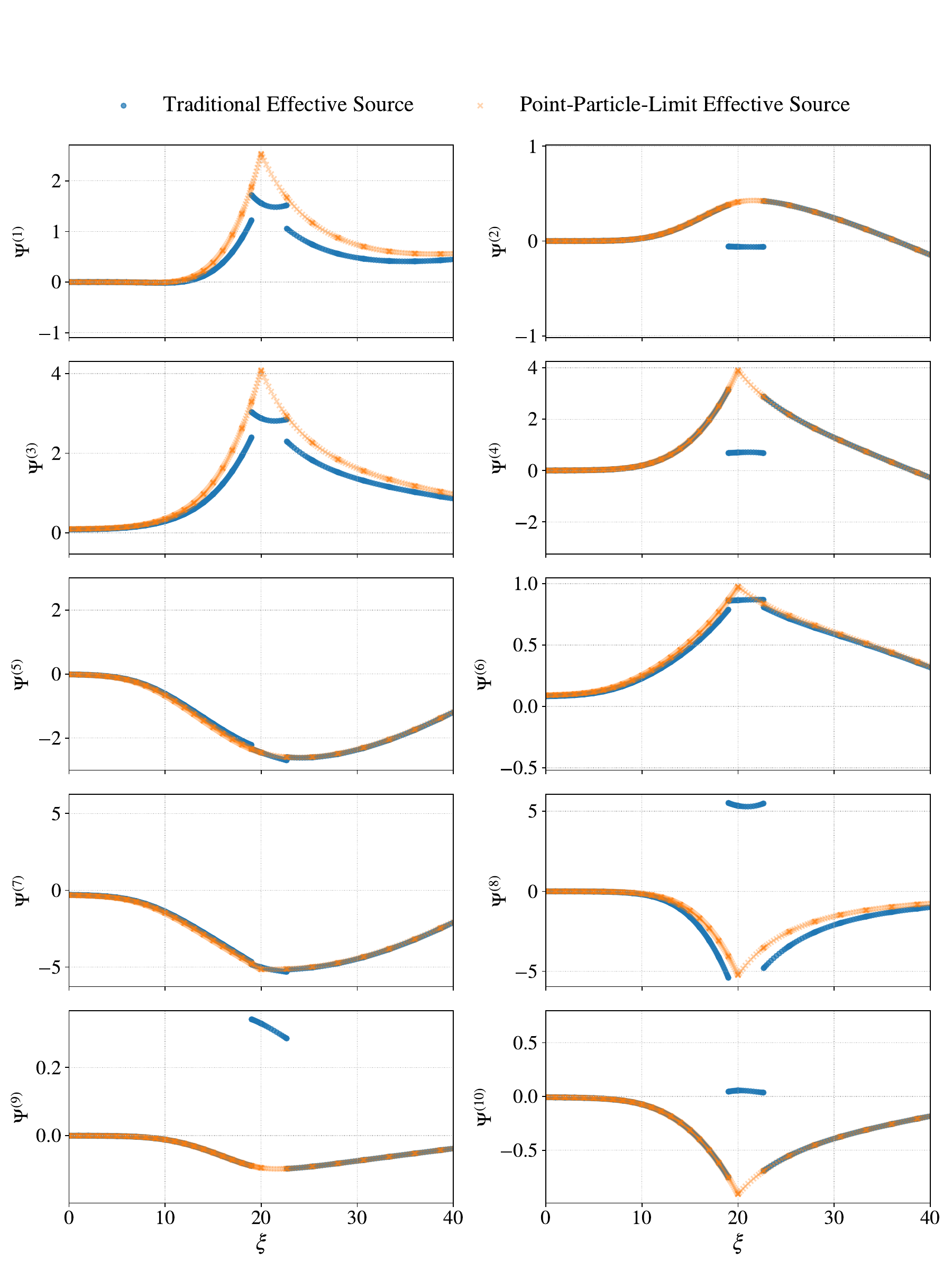}\\[8pt]
\parbox[c]{15.0cm}{\footnotesize{\bf Fig.~\thefigure.} Radial profile of the real components of the retarded metric field $\bar{h}^{(i)}_{\rm ret}=\Psi^{(i)}$ at the time slice $\lambda=1500$ s. Blue points show results from the TES method, and orange points correspond to the PPLES method. The retarded metric fields are compared across the computational domain, with the particle located at $\xi=20$. The two methods show excellent agreement in the far-field region, while differences near the particle arise from distinct implementations.}
\label{Fig2:e0}
\end{center}
Physically, the two methods should yield identical retarded fields. However, the observed deviation is largest near the tube boundaries and decays gradually with radial distance, suggesting that a straightforward application of the TES method may produce incorrect retarded fields in the Lorenz gauge.

The origin of this discrepancy can be attributed to two factors: the use of vanishing initial data, which is unphysical in the presence of discontinuities, and the enforcement of jump conditions at the world-tube interface.
In the DG framework, numerical fluxes depend only on $\partial_t \bar{h}^{(i)}$ and $\partial_r \bar{h}^{(i)}$, while the jump in $\bar{h}^{(i)}$ itself is imposed only weakly in an integral sense rather than pointwise.
This weak enforcement results in incomplete control of interface behavior, giving rise to spurious oscillations and persistent offsets, particularly in coupled systems.
To mitigate the influence of the unphysical vanishing initial data and the jump conditions, a constraint mechanism is introduced to suppress the resulting non-physical drift.
This is achieved by appending damping terms $\kappa \Psi^{(i)}$, $\kappa \Pi^{(i)}$, and $\kappa \Phi^{(i)}$ to the evolution equations, which enforce exponential decay of any deviation of the averaged field from zero, with the damping coefficient ramped down to zero after a relaxation period so that the constraint terms vanish as the system evolves.
The damping coefficient is then ramped down linearly according to
\begin{equation}
\kappa = 0.02 \times \bigl(1 - \lambda/800\bigr), \qquad 0 < \lambda < 800,
\end{equation}
and vanishes for $\lambda \geq 800$, after which the evolution returns to the original undamped equations.
With the damping mechanism applied, the TES results now align closely with the PPLES values at the particle location, as shown in Fig.~\ref{Fig3:e0}.
Similarly, Figure~\ref{Fig4:e0} demonstrates good agreement between the retarded fields from the two methods outside the effective-source region.
In contrast to the systematic discrepancies observed earlier, the two solutions are now largely consistent.
Nevertheless, very small residual discrepancies persist, originating from a constant shift that cannot be fully eliminated under weak enforcement of jump conditions, as well as small time-dependent variations in the boundary jumps.
\begin{center}
\refstepcounter{figure}
\includegraphics[width=0.99\textwidth]{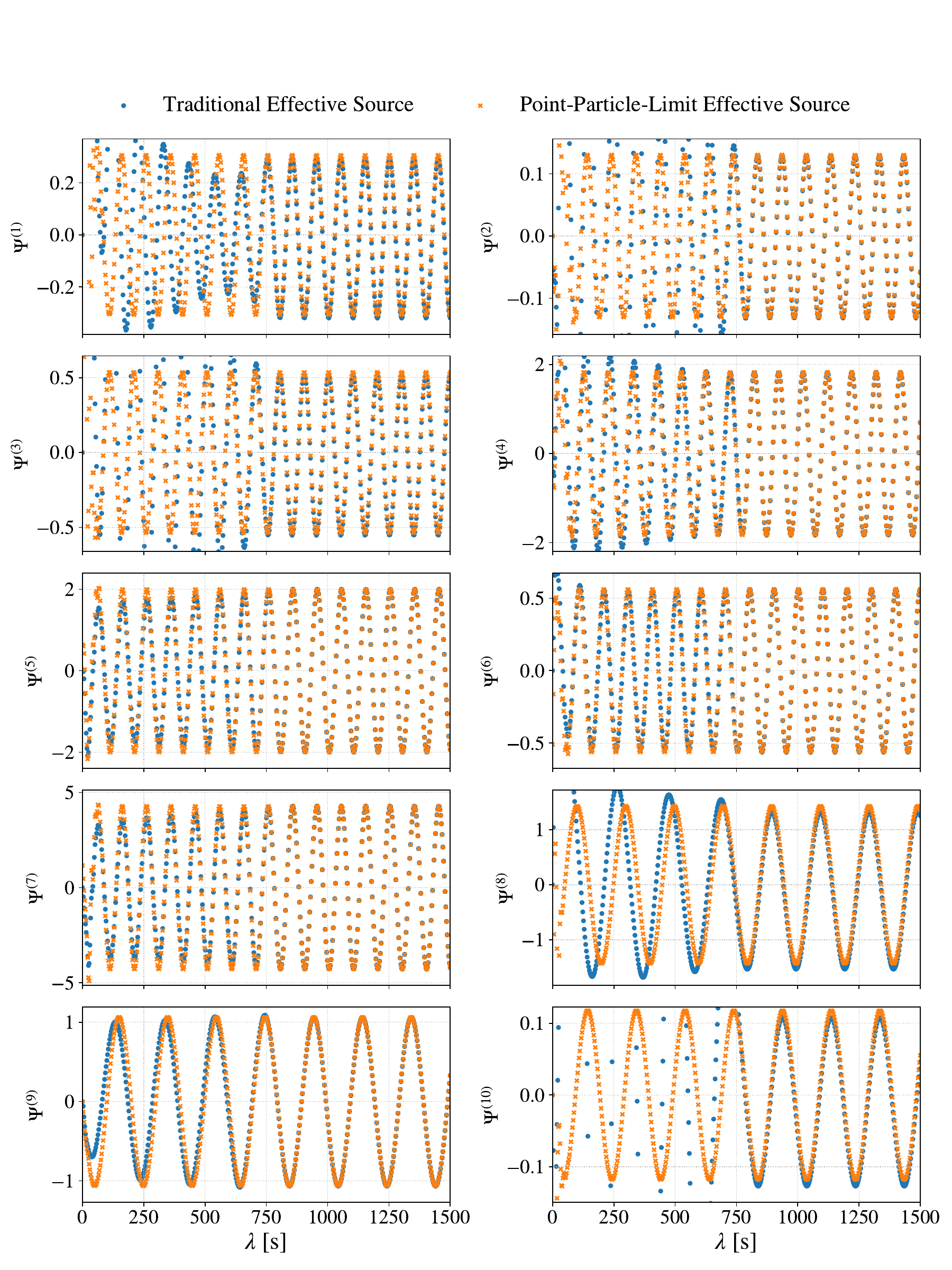}\\[8pt]
\parbox[c]{15.0cm}{\footnotesize{\bf Fig.~\thefigure.} Time evolution of the real components of the regular metric field $\bar{h}^{(i)}_R=\Psi^{(i)}$ at the particle position for a circular orbit around a Schwarzschild black hole with damping terms. Blue points show results from the TES method, and orange points correspond to the PPLES method. \CZ{Following the left-to-right, top-to-bottom order of the figure, the first seven display the even-parity components $\bar{h}^{(1-7)}_R$ for $(\ell,m)=(2,2)$, and the last three show the odd-parity components $\bar{h}^{(8-10)}_R$ for $(\ell,m)=(2,1)$.}}
\label{Fig3:e0}
\end{center}
\begin{center}
\refstepcounter{figure}
\includegraphics[width=1.0\textwidth]{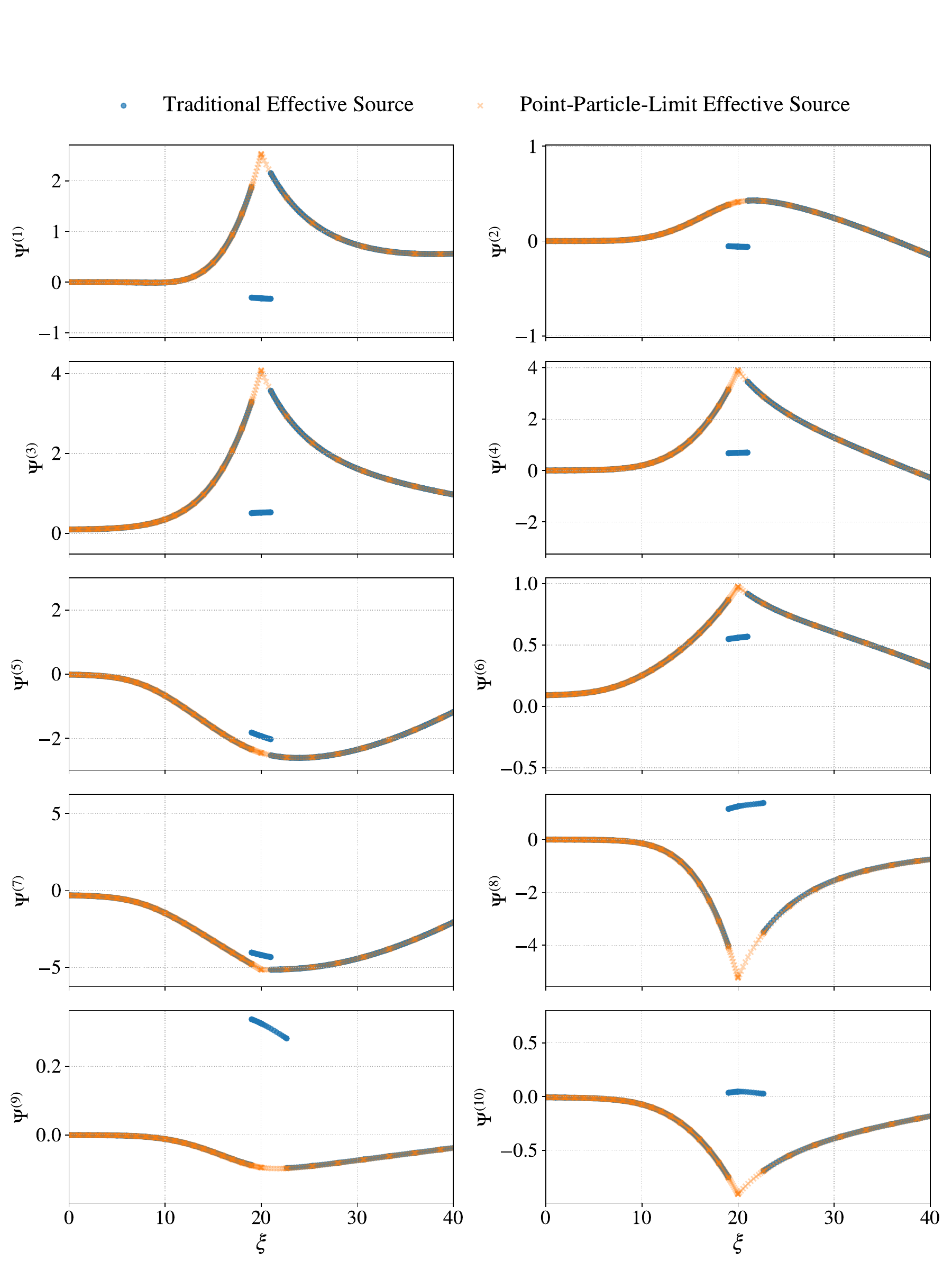}\\[8pt]
\parbox[c]{15.0cm}{\footnotesize{\bf Fig.~\thefigure.} Radial profile of the real components of the retarded metric field $\bar{h}^{(i)}_{\rm ret}=\Psi^{(i)}$ at the time slice $\lambda=1500$ s with damping terms. Blue points show results from the TES method, and orange points correspond to the PPLES method. The retarded metric fields are compared across the computational domain, with the particle located at $\xi=20$. The two methods show excellent agreement in the far-field region, together with the region near the particle.}
\label{Fig4:e0}
\end{center}
These results demonstrate that both the TES and PPLES methods yield identical energy and angular momentum fluxes.
This confirms the physical picture that the effective source influences the field only within its immediate region; far from this region, the field is governed solely by the boundary conditions and is independent of the method employed. Consequently, the retarded field outside the effective source is uniquely determined, consistent with the expectation that it does not depend on the details of the source representation. Moreover, as the effective source size approaches zero, the regular field $\bar{h}^{(i)}_{R}$ converges to a well-defined value.
Inside the effective source, the retarded field satisfies $\bar{h}^{(i)}_{\rm ret} = \bar{h}^{(i)}_{R} + \bar{h}^{(i)}_{S}$, and the PPLES method yields the same regular field as the TES method in this region.

Overall, these results confirm the validity of the PPLES approach while highlighting its advantages in simplicity and numerical robustness over the TES method.
For the PPLES method, the jump conditions are analytically known, imposed at a single point rather than at two boundaries, and free of the non-smooth behavior inherent to TES effective sources.
The simple analytic structure of the jumps also accelerates computation and reduces numerical error.
Moreover, since the jump in $\bar{h}^{(i)}_{\rm ret}$ at the particle location is zero, the numerical fluxes allow the constraints to be strictly satisfied during evolution, ensuring that the condition $\bar{h}^{(i)}_{\rm ret}=0$ holds throughout and that the issues arising from field jumps in the TES method are entirely avoided.
In terms of computational efficiency, the PPLES method offers a significant advantage as well that generating approximately three seconds of evolution requires about $600$ seconds per core with the TES method, whereas the PPLES method achieves the same evolution in roughly $30$ seconds$-$a reduction of about an order of magnitude.
Taken together, these advantages establish the PPLES method as a more robust, reliable, and efficient framework for computing Lorenz-gauge gravitational perturbations, particularly in systems governed by coupled field equations.

\subsection{Numerical Gravitational Wave Flux}
To validate our numerical implementation, we extract asymptotic fluxes directly from the Lorenz-gauge metric perturbation obtained numerically and compare them against results reported in the literature.
The energy fluxes of formulation proposed by Martel \cite{Martel:2003jj} and Poisson \cite{Poisson:2004cw} relies on the Regge-Wheeler and Zerilli-Moncrief perturbation functions $\Psi_{\rm RW}^{\ell m}$ and $\Psi_{\rm ZM}^{\ell m}$, which are connected to the Lorenz-gauge variables via
\begin{eqnarray*}\label{eq:RW}
\Psi_{\rm RW}^{lm} &=&
-\frac{(\ell-2)!}{2(\ell+2)!}\left[
\frac{\lambda}{r}\bar{h}^{(9)}
+ \frac{f}{r}\bar{h}^{(10)}
- \bar{h}^{(10)}_{,x}
\right], \\
\Psi_{\rm ZM}^{lm} &=& \label{eq:ZM}
\frac{2r}{\ell(\ell+1)(\lambda r + 6)}\left[
\bar{h}^{(1)} - \bar{h}^{(5)} - f \bar{h}^{(6)}
+ \frac{\ell(\ell+1)r+2}{2r} \bar{h}^{(3)}
- r \bar{h}^{(3)}_{,x}
+ \frac{\lambda r+6}{2\lambda r}\bar{h}^{(7)}
\right].
\end{eqnarray*}
 In terms of $\Psi_{\rm RW}^{\ell m}$ and
$\Psi_{\rm ZM}^{\ell m}$, the fluxes at infinity and horizon are given by
\cite{Martel:2003jj,Poisson:2004cw}
\begin{equation}
\langle \dot{E} \rangle_{\infty, H} =
\frac{1}{64\pi}\sum_{lm}\frac{(\ell+2)!}{(\ell-2)!}
\left\langle
4|\Psi_{\rm RW}^{\ell m}|^2
+|\dot{\Psi}_{\rm ZM}^{\ell m}|^2
\right\rangle, \label{eq:Edot-infinity}
\end{equation}
where the asterisk denotes complex conjugation, and the angle brackets $\left\langle\cdots\right\rangle$ indicate time average.
\begin{center}
\refstepcounter{table}
{\bf Table \thetable.} The energy flux of modes $(\ell,\ m)=(2,\ 1)$ and $(2,\ 2)$ in the gravitational waves absorbed by the black hole and radiated to infinity. Comparison with results from the frequency-domain approach in the BHPToolkit~\cite{BHPToolkit}.%
\label{table:compare_flux}
\vspace{2mm}
{\tabcolsep = 4.5mm
\begin{tabular}{lcccc}
\hline\hline
 & \multicolumn{2}{c}{Time Domain}  & \multicolumn{2}{c}{Frequency Domain} \\
\hline
\multicolumn{1}{l}{$r_p=10$} & $(2,\ 1)$& $(2,\ 2)$& $(2,\ 1)$& $(2,\ 2)$ \\
$\langle \dot{E} \rangle_{\infty}$
 & $9.675\times 10^{-8}$ &$2.684\times 10^{-5}$ & $9.658\times 10^{-8}$ &$2.684\times 10^{-5}$
   \\
 $\langle \dot{E} \rangle_{H}$
 & $6.134\times 10^{-10}$ &$5.654\times 10^{-9}$ & $6.134\times 10^{-10}$ &$5.654\times 10^{-9}$
   \\
\hline\hline
\end{tabular}}
\end{center}
Table~\ref{table:compare_flux} presents the energy fluxes $\langle \dot{E} \rangle_{\rm H}$ absorbed by the black hole and $\langle \dot{E} \rangle_{\rm \infty}$ radiated to infinity.
The close agreement of these computed fluxes confirms the validity of both our DG time-domain code and the PPLES method.

\subsection{Numerical Gravitational Self-Force}
With the Lorenz-gauge perturbation fields $\bar{h}^{(i)\ell m}$ in hand, the GSF components $F^{\alpha}$ can now be constructed.
Following \cite{Barack:2007tm,Barack:2010tm}, the GSF components are defined as a tensor field at an arbitrary spacetime point  for a given worldline point particle position
\begin{equation}\label{Ffull}
F^{\alpha}=m_p\, k^{\alpha\beta\gamma\delta} \bar h_{\beta\gamma;\delta}
\end{equation}
where
\begin{equation}\label{k}
k^{\alpha\beta\gamma\delta}=
         g^{\alpha\delta}u^{\beta}u^{\gamma}/2
        -g^{\alpha\beta}u^{\gamma}u^{\delta}
        -u^{\alpha}u^{\beta}u^{\gamma}u^{\delta}/2
        +u^{\alpha}g^{\beta\gamma}u^{\delta}/4
        +g^{\alpha\delta}g^{\beta\gamma}/4,
\end{equation}
where $g^{\alpha\delta}$ is the background metric, and $u^{\alpha}$ are the values of the contravariant components of the four-velocity at the particle position. 
Expanding the metric perturbation in tensor harmonics, the GSF takes the form
\begin{eqnarray} \label{Ffull2}
F^{\alpha}(\theta,\varphi;r_p)
&=&\frac{m_p^2}{r_p^2}\sum_{\ell=0}^{\infty}\sum_{m=-\ell}^{\ell}
\left\{
f_{0 }^{\alpha \ell m} Y^{\ell m}+
f_{1 }^{\alpha \ell m} \sin^2\theta\, Y^{\ell m}+
f_{2 }^{\alpha \ell m} \cos\theta\sin\theta\, Y^{\ell m}_{,\theta}\right.\nonumber\\
& &
+\left.f_{3 }^{\alpha \ell m} \sin^2\theta\, Y^{\ell m}_{,\theta\theta}
+f_{4 }^{\alpha\ell m} (\cos\theta Y^{\ell m}-\sin\theta Y^{\ell m}_{,\theta})\right.\nonumber\\
& &
+\left.f_{5 }^{\alpha \ell m} \sin\theta\, Y^{\ell m}_{,\theta}+
f_{6 }^{\alpha \ell m} \sin^3\theta\, Y^{\ell m}_{,\theta}+
f_{7 }^{\alpha \ell m} \cos\theta\sin^2\theta\, Y^{\ell m}_{,\theta\theta}
\right\},
\end{eqnarray}
where the coefficients $f_{n }^{\alpha \ell m}$ are constructed from the fields $\bar h^{(i)\ell m}_R$ and their first derivatives evaluated at the particle position.
The explicit expressions for $f_{n}^{\alpha \ell m}$ are lengthy and can be found in the Appendix of Ref. \cite{Barack:2010tm}.

The temporal and radial components of the GSF were calculated for two different orbital radii, $r_p = 8$ and $r_p = 10$.
To ensure a relative error of at least $\sim 10^{-3}$,  the sum is truncated at $\ell_{\rm max}=15$. 
The contributions from different $\ell$ modes to the GSF are shown in Table~\ref{table:compare-Edot}.
\begin{center}
\refstepcounter{table}
{\bf Table \thetable.} Mode-by-mode contributions to the temporal and radial components of the GSF for a point particle in a circular orbit around a Schwarzschild black hole, computed using the PPLES method with the sum truncated at $\ell_{\rm max}=15$. Results are shown for orbital radii $r_p = 8$ and $r_p = 10$. The total contributions from the present method are compared with those from Ref.~\cite{Barack:2007tm} at the same truncation, as well as with the high-accuracy reference values from Refs.~\cite{Barack:2007tm,Akcay:2010dx}. Relative differences between the present method and the high-accuracy reference values are reported, demonstrating agreement to within $\sim 10^{-4}$ for $F^t$ and $\sim 10^{-3}$ for $F^r$.
\label{table:compare-Edot}
\vspace{3mm}
\begin{tabular}{c|c|c|c|c}
\hline\hline
&\multicolumn{2}{c|}{$r_p=8$}&\multicolumn{2}{c}{$r_p=10$} \\
\hline
$(\ell, m)$ & $F^t$
       & $F^r$ & $F^t$
       & $F^r$ \\
\hline
$(0,0)$  & $0$ 
       & $-4.8632 \times 10^{-3}$ & $0$ 
       & $-2.1019 \times 10^{-3}$ \\
\hline
$(1,0)$   & $0$ 
       & $4.2953 \times 10^{-3}$ & $0$ 
       & $1.8911 \times 10^{-3}$\\
\hline
$(1,\pm 1)$   & $1.5503 \times 10^{-12}$ 
       & $1.9984 \times 10^{-2}$ 
       & $-1.0875 \times 10^{-13}$ 
       & $1.3959 \times 10^{-2}$\\
\hline
$(2, 0)$   & $4.8464 \times 10^{-9}$ 
       & $-2.6138 \times 10^{-4}$ & $-5.7143 \times 10^{-10}$ 
       & $-5.2752 \times 10^{-5}$\\
\hline
$(2,\pm 1)$   & $-1.3497 \times 10^{-6}$ 
       & $1.0436 \times 10^{-3}$ & $-2.8941 \times 10^{-7}$ 
       & $4.6090 \times 10^{-4}$\\
\hline
$(2,\pm 2)$   & $-2.7814 \times 10^{-4}$ 
       & $-1.5906 \times 10^{-3}$ & $-8.0228 \times 10^{-5}$ 
       & $-7.0637 \times 10^{-4}$\\
\hline
\multicolumn{5}{c}{...} \\
\hline
$\rm Total$ $(\ell_{\rm max}=15)$  
& $-3.3077 \times 10^{-4}$ 
       & $1.8358 \times 10^{-2}$ & $-9.1914 \times 10^{-5}$ 
       & $1.3384 \times 10^{-2}$\\
\hline
Ref. \cite{Barack:2007tm} $(\ell_{\rm max}=15)$   
       & $-3.3074 \times 10^{-4}$ 
       & $1.8532 \times 10^{-2}$ 
       & $-9.19067 \times 10^{-5}$ 
       & $1.3470\times 10^{-2}$\\
\hline
Refs. \cite{Barack:2007tm,Akcay:2010dx}   & $-3.3074 \times 10^{-4}$ 
       & $1.8357 \times 10^{-2}$ & $-9.1906 \times 10^{-5}$ 
       & $1.3389 \times 10^{-2}$\\
\hline
Rel. Diff.   & $9.0705 \times 10^{-5}$ 
       & $5.4475 \times 10^{-5}$ & $8.7045 \times 10^{-5}$ 
       & $3.7344 \times 10^{-4}$\\
\hline\hline
\end{tabular}
\end{center}
The computation error in $F^t$ is estimated at $\lesssim 10^{-4}$, while $F^r$ is accurate to within at least $\lesssim 10^{-3}$ for all radii considered, compared to the results from Refs. \cite{Barack:2007tm,Akcay:2010dx}.
The corresponding results from Ref.~\cite{Barack:2007tm} at the same truncation are shown in the table.
In comparison, the present method yields more accurate results at $\ell_{\rm max}=15$ than those obtained by Barack and Sago \cite{Barack:2007tm}, demonstrating improved precision under the same truncation.
Thus, our DG framework, together with the PPLES approach, ensures a more accurate calculation compared with the finite-difference scheme.

\section{Conclusions}\label{sec6}
In this paper, we present a novel PPLES approach for computing time-domain GSF in the Lorenz gauge, addressing critical limitations of TES methods that have hindered efficient, high-accuracy GSF calculations for EMRIs.
By analytically taking the effective source size to zero, the problem reduces to jump conditions at the particle position that align naturally with the DG framework, eliminating regularization parameters in the mode-sum method and worldtube boundary matching.
The PPLES method naturally integrates with the DG numerical scheme, whose inherent support for discontinuous solutions enables precise enforcement of the jump conditions through modified numerical fluxes at the particle interface.
We validate the PPLES approach using the canonical test case of a point particle in circular orbit around a Schwarzschild black hole, performing direct comparisons with the TES method implemented via the standard worldtube framework.
Quantitatively, the PPLES method achieves about one order reduction in computational runtime compared to the TES approach, and GSF components summed to $\ell_{\rm max}=15$ achieve relative differences below $10^{-4}$ (temporal) and $10^{-3}$ (radial) compared to established results,
outperforming finite-difference-based implementations for mode-sum method at the same multipole truncation.
The PPLES framework extends straightforwardly to eccentric geodesic orbits, where moving particle positions can be accommodated through the coordinate transformation technique already incorporated in our implementation, as well as to Kerr spacetime backgrounds, by generalizing the singular field jump conditions to rotating black hole spacetimes.
Most significantly, the PPLES formulation avoids the logarithmic mode divergences that complicate second-order GSF calculations in mode-sum schemes, making it particularly well-suited for the development of second-order GSF formalisms, enabling high-accuracy waveform templates for EMRIs for space-based detectors such as LISA, TianQin, and Taiji.

\addcontentsline{toc}{chapter}{Appendix A: Appendix section heading}
\section*{Appendix A: Basis of tensor harmonics} \label{AppA}
The tensor harmonics $Y^{(i)\ell m}_{\alpha\beta}$ are \cite{Barack:2005nr,Barack:2010tm}
\begin{equation}
Y^{(1)\ell m}_{\alpha\beta}=\frac{1}{\sqrt{2}}\left(
\begin{array}{c c c c}
1 & 0 & 0 & 0 \\
0 & f^{-2} & 0 & 0 \\
0 & 0 & 0 & 0 \\
0 & 0 & 0 & 0
\end{array}\right) Y^{\ell m},
\quad\quad
Y^{(2)\ell m}_{\alpha\beta}=\frac{f^{-1}}{\sqrt{2}}\left(
\begin{array}{c c c c}
0 & 1 & 0 & 0 \\
1 & 0 & 0 & 0 \\
0 & 0 & 0 & 0 \\
0 & 0 & 0 & 0
\end{array}\right) Y^{\ell m},
\end{equation}

\begin{equation} 
Y^{(3)\ell m}_{\alpha\beta}=\frac{f}{\sqrt{2}}\left(
\begin{array}{c c c c}
1 & 0 & 0 & 0 \\
0 & -f^{-2} & 0 & 0 \\
0 & 0 & 0 & 0 \\
0 & 0 & 0 & 0
\end{array}\right) Y^{\ell m},
\end{equation}

\begin{equation} 
Y^{(4)\ell m}_{\alpha\beta}=
\frac{r}{\sqrt{2 \ell( \ell+1)}}\left(
\begin{array}{c c c c}
0                 & 0 & \partial_{\theta} & \partial_{\varphi} \\
0                 & 0 &        0          &         0          \\
\partial_{\theta} & 0 &        0          &         0          \\
\partial_{\varphi}& 0 &        0          &         0
\end{array}\right) Y^{\ell m},
\quad\quad
Y^{(5)}_{\alpha\beta}=
\frac{rf^{-1}}{\sqrt{2 \ell( \ell+1)}}\left(
\begin{array}{c c c c}
0 &        0           &        0          &          0         \\
0 &        0           & \partial_{\theta} & \partial_{\varphi} \\
0 & \partial_{\theta}  &        0          &          0         \\
0 & \partial_{\varphi} &        0          &          0
\end{array}\right) Y^{\ell m},
\end{equation}

\begin{equation} 
Y^{(6)\ell m}_{\alpha\beta}=
\frac{r^2}{\sqrt{2}}\left(
\begin{array}{c c c c}
0 & 0 & 0 & 0 \\
0 & 0 & 0 & 0 \\
0 & 0 & 1 & 0 \\
0 & 0 & 0 & s^2
\end{array}\right) Y^{\ell m},
\quad\quad
Y^{(7)\ell m}_{\alpha\beta}=
\frac{r^2}{\sqrt{2\lambda  \ell( \ell+1)}}\left(
\begin{array}{c c c c}
0 & 0 & 0   & 0        \\
0 & 0 & 0   & 0        \\
0 & 0 & D_2 & D_1       \\
0 & 0 & D_1 & -s^2 D_2
\end{array}\right) Y^{\ell m},
\end{equation}

\begin{equation} 
Y^{(8)\ell m}_{\alpha\beta}=\frac{r}{\sqrt{2\ell(\ell+1)}}\left(
\begin{array}{c c c c}
0 &        0            & s^{-1}\partial_{\varphi} & -s\,\partial_{\theta} \\
0 &        0            &          0               &                       \\
s^{-1}\partial_{\varphi}&          0               &     0    &    0       \\
-s\,\partial_{\theta}   &          0               &     0    &    0
\end{array}\right) Y^{\ell m},
\end{equation}
\begin{equation} 
Y^{(9)\ell m}_{\alpha\beta}=\frac{rf^{-1}}{\sqrt{2\ell(\ell+1)}}\left(
\begin{array}{c c c c}
0 &        0                &        0                &           0         \\
0 &        0                & s^{-1}\partial_{\varphi} & -s\,\partial_{\theta} \\
0 & s^{-1}\partial_{\varphi}&        0                &          0          \\
0 & -s\,\partial_{\theta}   &        0                &          0
\end{array}\right) Y^{\ell m},
\end{equation}
\begin{equation} 
Y^{(10)\ell m}_{\alpha\beta}=
\frac{r^2}{\sqrt{2\lambda \ell(\ell+1)}}\left(
\begin{array}{c c c c}
0 & 0 & 0          & 0            \\
0 & 0 & 0          & 0            \\
0 & 0 & s^{-1}D_1 & -s\,D_2       \\
0 & 0 & -s\,D_2     & -s\,D_1
\end{array}\right) Y^{\ell m},
\end{equation}
where $D_1 \equiv 2(\partial_{\theta}-\cot\theta)\partial_{\varphi}$ and $D_2\equiv \partial_{\theta\theta}-\cot\theta\,\partial_{\theta}
-s^{-2} \partial_{\varphi\varphi}$ are the angular operators,  $Y^{\ell m}(\theta,\varphi)$ are the standard scalar spherical harmonics,  $\lambda \equiv (\ell-1)(\ell+2)$ and $s\equiv\sin\theta$.

\section*{Appendix B: The field equations}\label{AppB}
The terms ${\cal M}^{(i)}_{\;(j)}\bar h^{(j)}$ are \cite{Barack:2005nr}
\begin{align}
	\mathcal{M}^{(1)}{}_{(j)} \hb{j} & =  \f{f}{r^2} \hb{3}_{,r_\ast} + \f{f}{2r^2}\left(1-\f{4}{r}\right) \left(\hb{1}-\hb{5} - f \hb{3} \right) - \f{f^2}{2r^2} 	\left(1-\f{6}{r}\right) \hb{6}, \label{eq:eq_R1} \\
	\mathcal{M}^{(2)}{}_{(j)} \hb{j} & =  \f{1}{2}ff' \hb{3}_{,r_\ast} + \f{1}{2} f'  \left[ \hb{2}_{,t}-\hb{1}_{,t} + \hb{2}_{,r_\ast} - \hb{1}_{,r_\ast} \right] + 	\f{f^2}{2 r^2} \left(\hb{2}-\hb{4}\right) \\ & - \f{f f'}{2r}\left( \hb{1} - \hb{5} - f \hb{3} - 2f \hb{6} \right), \label{eq:eq_R2} \\
	\mathcal{M}^{(3)}{}_{(j)} \hb{j} & =  - \f{f}{2r^2} \left[\hb{1} - \hb{5} - \left(1-\f{4}{r}\right) \left(\hb{3} + \hb{6}\right) \right], \label{eq:eq_R3} \\
	\mathcal{M}^{(4)}{}_{(j)} \hb{j} & =  \f{1}{4} f' \left[ \hb{4}_{,t}-\hb{5}_{,t} + \hb{4}_{,r_\ast} - \hb{5}_{,r_\ast} \right] - \f{1}{2} \ell (\ell+1) \f{f}{ r^2} 	\hb{2}   \\ & - \f{f f'}{4r}\left( 3\hb{4} + 2\hb{5} - \hb{7} + \ell (\ell+1) \hb{6} \right)  , \label{eq:eq_R4} \\
	\mathcal{M}^{(5)}{}_{(j)} \hb{j} & =  \f{f}{r^2} \left[ \left(1-\f{9}{2r}\right) \hb{5} - \f{\ell(\ell+1)}{2}\left(\hb{1} - f \hb{3} \right)\right.\\
	&\left.+ \f{1}{2} \left(1-\f{3}{r}\right) 	\left( \ell(\ell+1) \hb{6} - \hb{7} \right) \right], \label{eq:eq_R5} \\
	\mathcal{M}^{(6)}{}_{(j)} \hb{j} & =  - \f{f}{2r^2} \left[\hb{1} - \hb{5} - \left(1-\f{4}{r}\right) \left(\hb{3} + \hb{6}\right) \right], \label{eq:eq_R6} \\
	\mathcal{M}^{(7)}{}_{(j)} \hb{j} & =  - \f{f}{2r^2} \left(\hb{7} + \lambda \hb{5} \right), \label{eq:eq_R7} \\
	\mathcal{M}^{(8)}{}_{(j)} \hb{j} & = \f{1}{4} f' \left[\hb{8}_{,t}-\hb{9}_{,t} + \hb{8}_{,r_\ast} - \hb{9}_{,r_\ast} \right] -  \f{f f'}{4r}\left( 3\hb{8} + 	2\hb{9} - \hb{10}  \right) , \label{eq:eq_R8} \\
	\mathcal{M}^{(9)}{}_{(j)} \hb{j} & =  \f{f}{r^2}\left(1-\f{9}{2r}\right) \hb{9} - \f{f}{2r^2}\left(1-\f{3}{r}\right) \hb{10}, \label{eq:eq_R9} \\  
	\mathcal{M}^{(10)}{}_{(j)}\hb{j} &=  -\frac{f}{2r^2}\left(\hb{10} + \lambda\hb{9}\right).	  \label{eq:eq_R10}
\end{align}

\addcontentsline{toc}{chapter}{Acknowledgment}
\section*{Acknowledgment}
This work is supported in part by the National Natural Science Foundation of China under Grant Nos. 12505076, 12535002 and 12588101.


\providecommand{\newblock}{}

\end{document}